# On the possibility to consider fullerene shell $C_{60}$ as a conducting sphere

M. Ya. Amusia[1, 2] and A. S. Baltenkov[3]

[1]Racah Institute of Physics, The Hebrew University, Jerusalem 91904, Israel
[2]Ioffe Physical-Technical Institute, St.-Petersburg 194021, Russia
[3]Arifov Institute of Electronics, Tashkent, 700125, Uzbekistan

**Abstract**

Correctness of the model representing the fullerene shell $C_{60}$ as a conducting sphere has been analyzed. The static and dynamical polarizabilities of the molecule $C_{60}$ have been calculated on the basis of experimental data on the photoabsorption cross-section of fullerene. It has been shown that the real $C_{60}$ in the static electric field behaves most likely as a set of separate carbon atoms rather than as a conducting sphere and its static polarizability exceeds by more than two times that of conducting sphere.

1. The aim of this Letter is to demonstrate that by studying photoeffect from mesoscopic objects one can conclude to which extent they are a "conductor" or "dielectric". These terms characterize the ability of the object to conduct electricity and are applicable, strictly speaking, only to macroscopic bodies. Therefore, it is meaningless to speak about conductivity of electrons that belong to such a microscopic object as a single, even a multi-electron, atom. However, for mesoscopic objects that include many atoms the feature of conductivity can become meaningful starting from a given number of constituent atoms. This makes meaningful the question whether mesoscopic system behaves as a conductor, i.e. whether its electrons can move freely under the action of an external electric field.

Since it is impossible to make a mesoscopic system an element of a normal electric circuit, other features characteristic of a conductor must be used to answer a question whether this system behaves like a conductor or not. As such, we use a quite natural definition that a static external electric field cannot penetrate inside a conductiong body. As an object of investigation and as a concrete example of a mesoscopic system we choose in this Letter the fullerene $C_{60}$ molecule. The effective electric field $\mathbf{E}_{eff}(\omega)$ that is at the fullerene center when external electric field $\mathbf{E}(\omega)$ with frequency $\omega$ is applied to it will be calculated. This modification of the external field comes from the effect of the dynamic dipole polarizability $\alpha_d(\omega)$ of the fullerene that can be expressed via the total photoionization cross section $\sigma(\omega)$ of $C_{60}$. Then the ratio $\eta(\omega) \equiv |\mathbf{E}_{eff}(\omega)/\mathbf{E}(\omega)|$ will be calculated to find out whether $\eta(0)$ is zero. It proved to be that this ratio is $\eta(0) \approx 1.5$ and thus $C_{60}$, contrary to the conclusion in Ref. [1], is strongly non-metallic. It is demonstrated that it cannot screen, but only enhances the external field inside $C_{60}$.

2. The experimental studies of fullerene interaction with electromagnetic radiation are evidence of a high degree of collectivization of the $2s^2$ and $2p^2$ electrons of the carbon atoms forming a fullerene. This manifest itself, first of all, in existing the Giant Resonance in the cross section of the $C_{60}$ fullerene photoabsorption that was discovered and then investigated in a number of experimental studies [2, 3]. Virtual excitation of these electrons significantly influences the radiative and Auger - decay processes taking place in so-called endohedral atoms A@$C_{60}$, i.e. in the atoms located inside the fullerene cage [4, 5]. Plasmon excitations in an ensemble of delocalized electrons of $C_{60}$ also influence radically the processes of photoabsorption of confined atoms [6].

In Ref. [1] it was considered to which extent the external electromagnetic field could reach the confined atom in A@$C_{60}$. The aim was to find out the possibilities of photoelectron spectroscopy to study the confined atoms and determine the frequency windows, at which these atoms can be effectively excited. To do this, the collective behavior of the $C_{60}$-electrons was used as a ground to replace the real fullerene cage by a spherical metallic container, at the center of which the ionized atom is located. Let us analyze whether such a model description of $C_{60}$ screening of an encapsulated A atom is realistic and whether the fullerene shell really resembles a hollow metal sphere.

It was shown in Ref. [1] by the methods of classical electrodynamics and in Refs. [5, 6] within the framework of quantum-mechanical consideration that the photoionization cross section of endohedral atoms with taking into account the dipole collective excitations of $C_{60}$ electrons is a product of two dynamical factors: the photoionization cross section of the free A atom and the function $F(\omega)$ depending on photon energy. This function has the following form

$$F(\omega) = \left|1 - \frac{\alpha_d(\omega)}{R^3}\right|^2. \qquad (1)$$

Here $\alpha_d(\omega)$ is the dynamical dipole polarizability of $C_{60}$ fullerene and $R$ is its radius. The reason for such factorization of the photoionization cross section of endohedral atoms is that the radius of the fullerene shell $R \approx 6.72$ significantly exceeds the radii of electronic shells of the encapsulated atom[*/].

The function Eq. (1) is connected with the ratio of the electric fields $\eta(\omega)$ at the fullerene center and outside the $C_{60}$ shell by the relation $F(\omega) = \eta^2(\omega)$. It is known that the static polarizability of a conducting sphere is $\alpha_d(0) = R^3$ [7]. Therefore, in the model of fullerene shell Ref. [1] the function Eq. (1) goes to zero for the limit of zero frequencies of radiation: $F(0) = 0$. This is a reflection of the fact that the static electric field $\mathbf{E}_{eff}(0)$ does not penetrate inside the conducting sphere ("Maxwell cage") and for this reason has no effect on the atom being inside. Consequently, within the model of conducting fullerene shell the encapsulated atom is not polarized by the external static electric field. According to Ref [1], $F(\omega)$ is small enough also in a relatively broad frequency region.

Generally speaking, there are no grounds to consider that the polarizability of the empty fullerene $C_{60}$ and that of the same molecule but containing inside an atom A are equal. Hence the assumption that the fullerene shell in the electric field behaves as a conducting sphere should be justified. The difference of the real static polarizability $\alpha_d(0)$ from $R^3$ defines the degree of metallicity of the fullerene shell and, consequently, the difference between the polarizability of the $C_{60}$ and A@$C_{60}$ molecules.

---

[*] The atomic system of units: $e = m = \hbar = 1$ is used throughout this paper.

The dynamical polarizability of $C_{60}$, as shown in Ref. [8], can be calculated from the experimental data on the photoabsorbtion of this molecule [2, 3]. Such an approach, unlike the method in Ref. [1], allows avoiding a number of the arbitrary assumptions inevitable for calculation of dynamic dipole polarizability of the metallic sphere by the methods of classical electrodynamics. An important circumstance to apply method [8] is that the photoionization cross-section of the fullerene $C_{60}$, $\sigma(\omega)$, is characterized by a Giant Resonance with frequency $\omega_{GR} \approx 22$ eV. The calculation shows that the oscillator strengths corresponding to the electron transition to the continuum are so that the electron transitions to the discrete spectrum can be neglected. This allows calculating the real part of the dynamic dipole polarizability of fullerene with the help of the dispersion relation keeping in them just the integral corresponding to the transitions of optical electron to the continuum:

$$\operatorname{Re}\alpha_d(\omega) = \frac{c}{2\pi^2} \int_I^\infty \frac{\sigma(\omega')d\omega'}{\omega'^2 - \omega^2}. \qquad (2)$$

Here $c$ is the speed of light, $I$ is the ionization potential of $C_{60}$. Exactly by the real part of polarizability Eq. (2) the static polarizability is defined because the imaginary part in the limit $\omega \to 0$ is equal to zero: $\operatorname{Im}\alpha_d(0) \equiv 0$. Hence for the static polarizability of $C_{60}$ from Eq. (2) we have the following expression

$$\alpha_d(0) = \frac{c}{2\pi^2} \int_I^\infty \frac{\sigma(\omega')d\omega'}{\omega'^2}. \qquad (3)$$

For the opposite limit of high frequencies of radiation $\omega \to \infty$ we obtain

$$\operatorname{Re}\alpha_d(\omega \to \infty) = -\frac{1}{\omega^2}\frac{c}{2\pi^2}\int_I^\infty \sigma(\omega')d\omega' = -\frac{N_{eff}}{\omega^2}. \qquad (4)$$

This limit transition can be used to control the correctness of the numerical results obtained with integrals (2) and (3). The calculation of the integral (4) with the experimental data for $\sigma(\omega)$ performed in Refs. [2, 3] shows that the total number of electrons $N_{eff}$ that take part in forming of the Giant Resonance in photoionization cross section is equal to $N_{eff} \approx 250$. This value differs only by ~ 4% from the number of collectivized $2s^2$ and $2p^2$ electrons of the 60 carbon atoms, each giving 4 electrons forming a system of 240 collectivized electrons.

3. The calculation results of the real and imaginary parts of dynamic polarizability of the fullerene shell $\alpha_d(\omega)$ are depicted in Fig.1. The experimental photoabsorption cross-section for $C_{60}$ taken from [3] is presented in the inset of this figure. As seen from this figure, the cross section is small at threshold (which also means the low intensity of discrete excitations) and has the form of a huge maximum - Giant Resonance - well above the threshold.

The frequency dependence of the imaginary part $\operatorname{Im}\alpha_d(\omega) = c\sigma(\omega)/4\pi\omega$, as it should be, is similar to the frequency dependence of $\sigma(\omega)$. A small peak in the cross section for photon energy $\omega \approx 5$ eV is transformed into a significant maximum, which is explained by a small value of photon energy as compared to the energy of the Giant resonance $\omega_{GR} \approx 22\,\text{eV}$. The real part of the polarizability obtained as a result of integration according to formula (2) behaves more systematically and with the raise of radiation frequency $\omega$, as it should be, decreases according to Eq. (4).

As is seen from Fig. 1, the static polarizability of $C_{60}$ is equal to $\alpha_d(0) = 760$ while for the model of conducting sphere it is equal to $\alpha_d^m(0) = 303$. This value is more than two times smaller than the value obtained with the experimental photoabsorption data [2, 3]. Note that the static polarizability of a free C atom is equal to $\alpha_d^C(0) = 14$ [9]. Therefore, by the order of a magnitude, the polarizability of 60 carbon atoms is the sum of their polarizabilities: $\alpha_d^{60C}(0) = 14 \times 60 = 840$. This value is remarkably close to experimental one $\alpha_d(0) = 760$.

The function $\eta(\omega)$ defining the ratio of the values of electric fields at the center of the $C_{60}$ shell and outside is given in Fig. 2. In the static limit this ratio is equal to $\eta(0) \approx 1.5$. The reason for such strengthening of the field inside the molecule is evidently the dipole polarization of the carbon atoms, the electric fields of which enhance the external field.

4. The calculations performed allow the following conclusions. Despite the high degree of $2s^2$ and $2p^2$ electron collectivization manifesting itself as the Giant Resonance in the photoionization cross section $\sigma(\omega)$, the fullerene shell in the static electric field behaves most likely as a set of separate carbon atoms rather than as a conducting sphere. The static electric field at the center of $C_{60}$ is not equal to zero, but one and a half times stronger than the external electric field, contrary to what follows from the "metal sphere" model developed in Ref. [1]. The polarizability of the empty fullerene cage is incapable, as shown above, totally screen the external electric field and differs considerably from the polarizability of endohedral system A@$C_{60}$.

It is remarkable that the function $\eta(\omega)$ has quite a complex structure thus demonstrating rather tricky modification of the external field due to $C_{60}$ effect.

The authors are grateful for financial support to Bi-national Science Foundation, Grant 2002-064 and Israeli Science Foundation, Grant 174/03. This work was also supported by Uzbekistan National Foundation, Grant Ф-2-1-12.

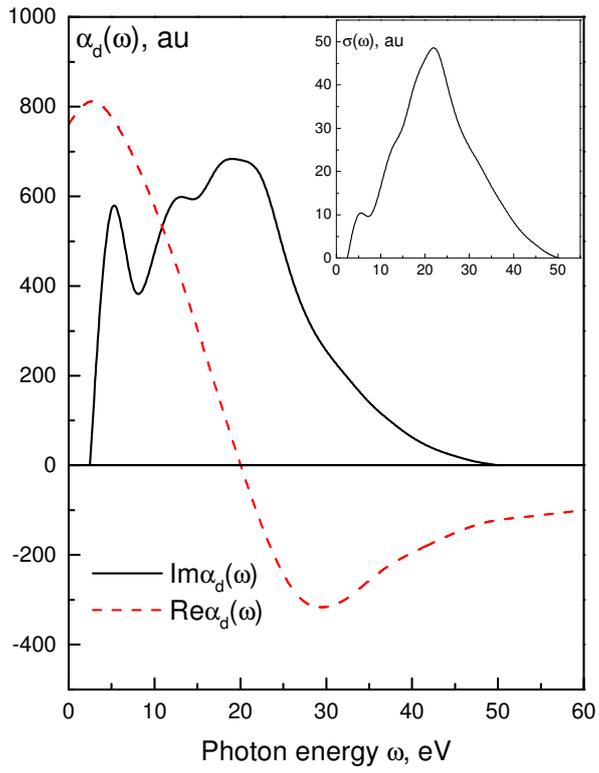

Fig. 1. Real and imagine parts of the dynamical polarizability of $C_{60}$.

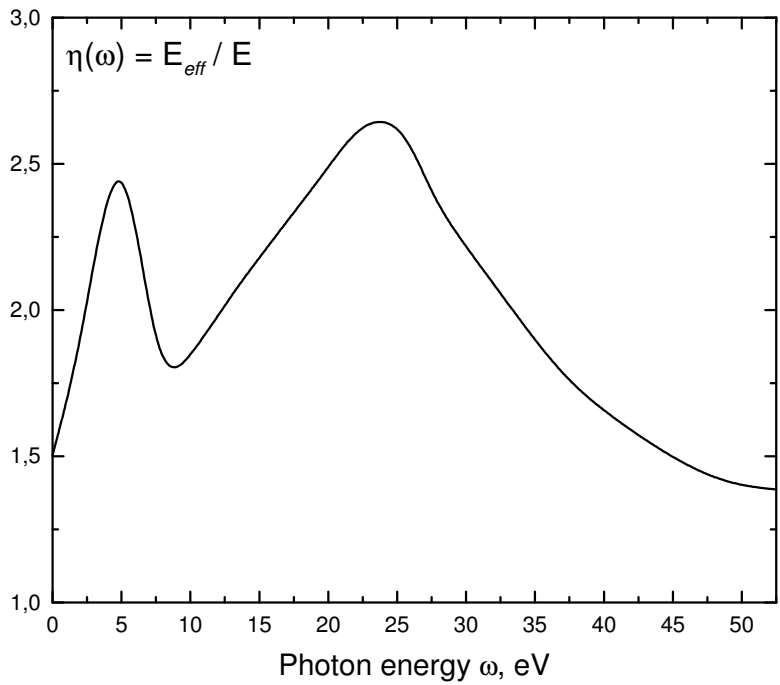

Fig. 2. The ratio of values of electric fields as a function of photon energy.